\documentclass[aps,prl,preprint,groupedaddress,showpacs]{revtex4}

\usepackage{graphics}
\usepackage{dcolumn}
\usepackage{bm}

\begin{document}


\title{ Revisiting the Perfect Lens with Loss}


\author{Weiguo Yang}
\email[]{wyang@email.wcu.edu}
\affiliation{ Department of Engineering \& Technology, Western Carolina University, Cullowhee, NC 28723, USA}

\author{ John O. Schenk and Michael A. Fiddy}
\email[]{mafiddy@uncc.edu}
\affiliation{Center for Optoelectronics and Optical Communications, University of North Carolina at Charlotte, NC 28223, USA}


\date{\today}

\begin{abstract}
We reconsider the refraction of evanescent waves at an interface between air and negative index medium under the assumption that negative index medium is necessarily dispersive and lossy.  We show that all evanescent waves in air will be refracted into decaying propagating waves inside a negative index medium, with different spatial frequency components having different propagation directions which are separated both in time and space; hence no refocus of these evanescent waves is possible.  Accordingly, all information encoded by evanescent waves will be lost in the image making sub-diffraction-limited imaging impossible. 
\end{abstract}


\maketitle


\section{Introduction}


Pendry\textquoteright s seminal paper on perfect lens \cite{pendry2000}, after some controversies \cite{ctr2,ctr3,ctr4,ctr5,ctr6,ctr7,ctr8,ctr9}, modifications \cite{pendryMod} and experimental verifications notably by Liu, et. al. \cite{Liu6} and Fang, et. al.\cite{FangScience}, established the surprising result that a slab of negative index material (NIM) amplifies evanescent waves, sustaining them through an NIM slab.  This therefore enables perfect imaging in the ideal case of no loss and sub-diffraction-limited imaging when the NIM is lossy \cite{pendry2000,pendryMod,FangScience}.  Since these evanescent waves carry high spatial frequency information about an object, this theory opens up the opportunity of realizing a higher resolution lens, perfect in the ideal case \cite{Science7}.  Recently, we have pointed out a possible inconsistency in Pendry\textquoteright s theory where physically sound assumptions have led to a set of self-contradictory equations in the limit as $\epsilon$ and $\mu$ tend to -1 \cite{yangaXiv}.  In order to resolve this paradox, different forms of solutions to wave equations are applied to positive and negative index materials which are both naturally assumed to obey the same Maxwell\textquoteright s equations in the frequency domain.  As a result, such solutions must maintain continuity of the reflection and transmission coefficients, i.e. continuity of the tangential component of the $k$ vector (i.e., $k_x$) approaching the propagation constant $k_0=(\epsilon\mu)^{1/2}\omega/c$ from either $k_x-k_0<0$ or $k_x-k_0>0$. \cite{yangaXiv}. We also assert that evanescent waves, since they do not transport energy and do not propagate, do not possess momentum and so momentum conservation arguments are ambiguous at best when modeling evanescent waves \cite{pendryPrivate}.

	In this paper, we will reconsider the refraction of evanescent waves at an interface of air and NIM under the assumption that NIM is necessarily dispersive and lossy \cite{stockmanPRL07}. We show that under this assumption all evanescent waves in air will be refracted into decaying propagation waves inside the negative index medium, with different spatial frequency components having different propagation directions which are separated both in time and space and, accordingly, no refocusing of these evanescent waves by the NIM slab to the image is possible.  As a result, all evanescent wave information is lost to the image and no perfect lens or sub-diffraction-limited imaging can be expected.  In light of this conclusion, we maintain that previous experimental verifications of Pendry\textquoteright s theory for sub-diffraction-limited optical imaging (e.g. Ref. \cite{FangScience}) can only be explained by other theories such as the coupling of surface plasmonic states \cite{yangaXiv,rao03}, should metallic structures be involved.

\section{ Refraction at air and NIM interface}
Assuming that the interface between air ($z<0$) and NIM ($z>0$) is parallel to $x$ axis and the $(x-z)$ plane is the principal plane that contains the propagation vector $\mathbf{k}=(k_x,k_z)$ and a normal of the interface, when $k_x<|\mathbf{k}|$, $k_z$ is a real number and plane waves $e^{i\mathbf{k}\cdot\mathbf{r}-i\omega t}$ are propagating waves. Without loss of generality, we will limit our discussion to the S polarization.  The treatment of P polarization is similar and straightforward and the conclusion is the same.  We start with the input evanescent waves in air, whose electric field is given by,

\begin{equation}
E_{0S+}=[0,1,0]\exp(ik_zz+ik_xx-i\omega t),
\end{equation} 

where the wave vector

\begin{equation}
k_z=+i\sqrt{k_x^2-\omega^2c^{-2}}, \qquad \omega^2c^{-2}<k_x^2.
\end{equation} 

The electric field of the reflected light, following Pendry\textquoteright s notation\cite{pendry2000}, is given by,

\begin{equation}
E_{0S-}=r[0,1,0]\exp(-ik_zz+ik_xx-i\omega t),
\end{equation} 

where $r$ is the reflection coefficient.  The transmitted electric field is given by,

\begin{equation}
E_{1S+}=t[0,1,0]\exp(ik^\prime_zz+ik^\prime_xx-i\omega t),
\end{equation}

where

\begin{equation}
k^{\prime2}_x+ k^{\prime2}_z=\epsilon\mu\omega^2c^{-2} =(\epsilon^\prime+i\epsilon^{\prime\prime})(\mu^\prime+i\mu^{\prime\prime})\omega^2c^{-2}
\label{Eq5}
\end{equation}

assuming the NIM is necessarily dispersive and lossy ($\epsilon^{\prime\prime}\neq 0$ and $\mu^{\prime\prime}\neq 0$). Note that for time dependence of the form $\exp(-i\omega t)$ chosen here, a passive/lossy material has positive imaginary parts for $\epsilon$ and $\mu$. 

The boundary conditions at the air and NIM interface are $E_{0y}=E_{1y}$, $B_{0z}=B_{1z}$, and $B_{0x}=B_{1x}/(\mu^\prime+i\mu^{\prime\prime})$, where $E$ and $B$ are the electric and magnetic fields, related by the Maxwell\textquoteright s equation $\nabla\times\mathbf{E}=i\omega\mathbf{B}$. Accordingly, one has at the interface, respectively,

\begin{equation}
\exp(ik_xx-i\omega t)+r\exp(ik_xx-i\omega t)=t\exp(ik^\prime_xx-i\omega t)
\label{Eq6a}
\end{equation}

\begin{equation}
k_x\exp(ik_xx-i\omega t)+rk_x\exp(ik_xx-i\omega t)=tk^\prime_x\exp(ik^\prime_xx-i\omega t)
\label{Eq6b}
\end{equation}

\begin{equation}
k_z\exp(ik_xx-i\omega t)-rk_z\exp(ik_xx-i\omega t)=tk^\prime_z\exp(ik^\prime_xx-i\omega t)/(\mu^\prime+i\mu^{\prime\prime})
\label{Eq6c}
\end{equation}

Obviously, if $k_z$ is not zero, then $t$ cannot be zero. Since if $t=0$, one has from Eq.\ref{Eq6a} $1+r=0$, or $r=-1$. But this results in $2k_z=0$ from Eq.\ref{Eq6c}, which conflicts with the condition that $k_z$ is not zero. 

When $t$ is not equal to zero, one has $k_x=k^\prime_x$ (Snell\textquoteright s law of refraction), which is real number. Consequently from Eq.\ref{Eq5},

\begin{equation}
k^{\prime2}_z =(\epsilon^\prime+i\epsilon^{\prime\prime})(\mu^\prime+i\mu^{\prime\prime})\omega^2c^{-2}- k_x^2
\label{Eq7}
\end{equation}

From causality arguments, the transmitted wave has to decay away from the interface. Accordingly $\Im(k^\prime_z)>0$. For lossless medium in the limit $\epsilon\to-1$ and $\mu\to-1$, this requires $k^\prime_z=k_z$ \cite{pendry2000}. As the result, Eq.\ref{Eq6a} to Eq.\ref{Eq6c} become self-contradictory when input field is not zero, reducing to $1+r=t$ and $1-r=-t$ \cite{yangaXiv}. This difficulty can be avoided if $\epsilon$ or $\mu$ has non-vanishing imaginary part, for example, in lossy NIM.  Denote

\begin{equation}
k^\prime_z=\Re(k^\prime_z)+i\Im(k^\prime_z),
\label{Eq8}
\end{equation}

substitute Eq.\ref{Eq8} into Eq.\ref{Eq7}, one has for the imaginary part of the equation, 

\begin{equation}
2\Re(k^\prime_z)\Im(k^\prime_z)=(\epsilon^\prime\mu^{\prime\prime}+\epsilon^{\prime\prime}\mu^\prime)\omega^2c^{-2},
\label{Eq9}
\end{equation}

for $\epsilon^{\prime\prime}\neq 0$ and $\mu^{\prime\prime}\neq 0$, one has $\Re(k^\prime_z)\Im(k^\prime_z)\neq 0$ unless the real parts of $\epsilon$ and $\mu$ are both zero.  When the real parts of $\epsilon$ and $\mu$ are not simultaneously zero, real and imaginary part of $k^\prime_z$ are both non-zero and, as the result, the evanescent waves will be transformed inside the NIM decaying propagating waves. Eq.\ref{Eq7} also suggests that evanescent waves with different spatial frequency $k_x$ will have different $\Re(k^\prime_z)$ and, accordingly, be refracted into decaying propagating waves having different directions of propagation inside the NIM. Given the fixed thickness of an NIM slab, this will result in different optical phase accumulation for such waves having different spatial frequencies.  Since evanescent waves in air do not acquire optical phase along the $z$ direction, these components will reach the image plane behind the NIM slab with different phases.  As a result, the NIM slab does not refocus these evanescent waves to a perfect image and the information contained in the evanescent waves is most likely lost.  However, if the NIM is truly lossless or exhibits increased transparency through parametric amplification \cite{Natasha09} then we could contrive to make $\Im(k^\prime_z) \leq 0$. 

\section{Conclusion and Discussion}

In conclusion, we claim that, under the assumption that negative index medium is necessarily dispersive and lossy, all evanescent waves in air incident on a NIM boundary will be refracted into decaying propagating waves inside the negative index medium. Furthermore, evanescent waves with different spatial frequencies will also have different propagation directions inside the NIM and be separated both in time and space.  Accordingly, under these circumstances, the NIM slab does not refocus these evanescent waves to the image point. As a result, all information about the object encoded by evanescent waves will be lost in the image and sub-diffraction-limited imaging is impossible.  In light of this conclusion, we maintain that previous experimental verifications of Pendry\textquoteright s theory for sub-diffraction-limited optical imaging, for example Ref.\cite{FangScience}, can only be explained by other theories such as the coupling of surface plasmonic states.

\newpage
\begin{figure}
\resizebox{6in}{!}{\includegraphics{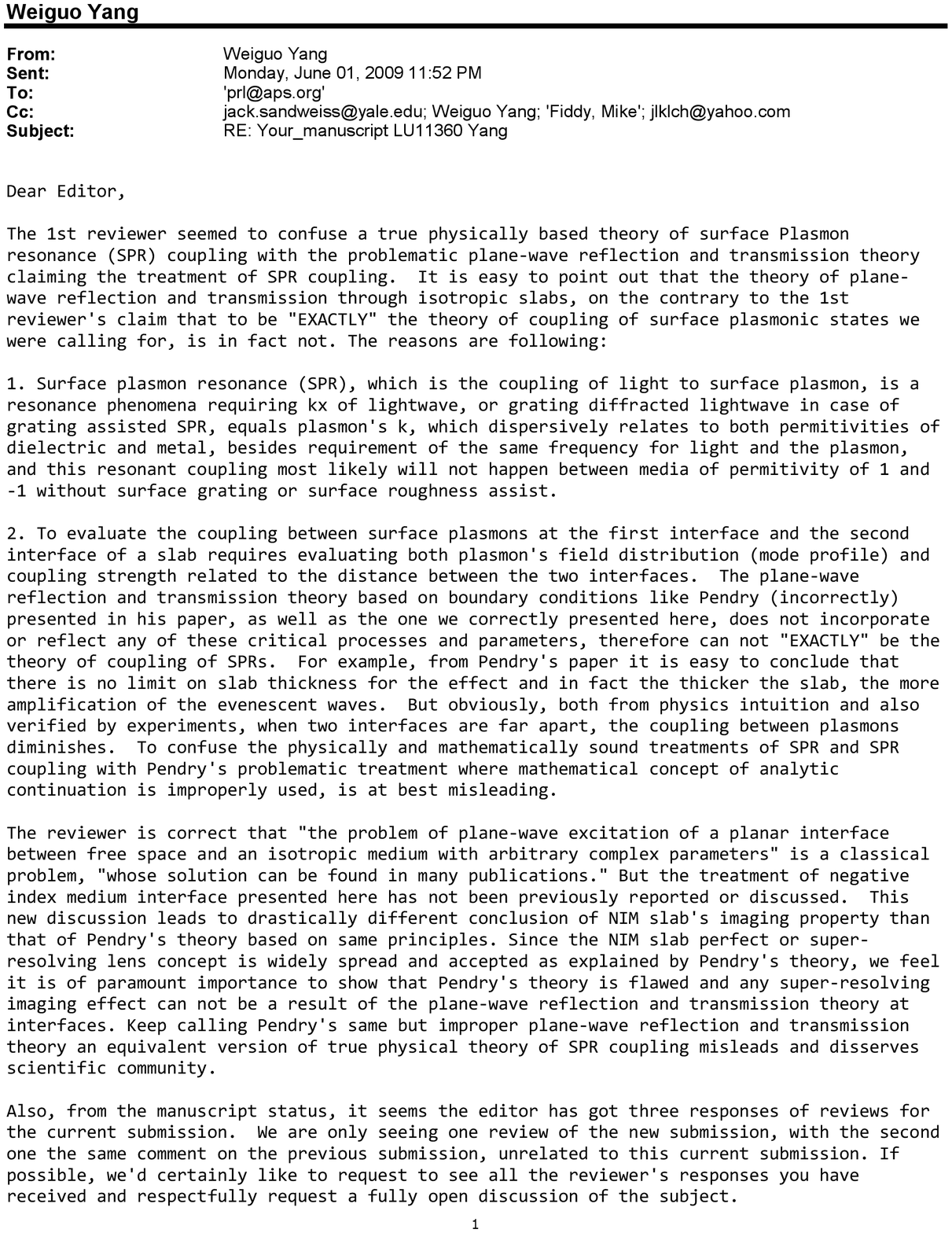}}%
\caption{Reply to PRL reviews. Reviews and Editor correspondences are omitted due to copyright restrictions.}
\end{figure}

\end{document}